\documentclass[final,1p,times]{elsarticle} 

\usepackage{graphicx}
\usepackage{amssymb} 
\usepackage{amsthm} 
\usepackage{lineno}

\newcommand{\bq}{\begin{equation}}
\newcommand{\eq}{\end{equation}}

\newcommand{\bqq}{\begin{eqnarray}}
\newcommand{\eqq}{\end{eqnarray}}
\newcommand{\PbPb}     {Pb--Pb}

\newcommand{\pt}{\ensuremath{p_{T}}}
\newcommand{\pT}{\ensuremath{p_{T}}}
\newcommand{\kt}{\ensuremath{k_{T}}}
\newcommand{\et}{\ensuremath{E_{\rm T}}}

\newcommand{\gev}     {\mbox{${\rm GeV}$}}
\newcommand{\tev}     {\mbox{${\rm TeV}$}}
\newcommand{\gmom}    {\mbox{${\rm GeV}/c$}}
\newcommand{\mmom}    {\mbox{${\rm MeV}/c$}}

\newcommand{\RAA}     {\mbox{$R_{\rm AA}$}}

\journal{Nuclear Physics A} 

\begin{document}

\begin{frontmatter} 

\title{Results on Jet Spectra and Structure from ALICE}

\author{Andreas Morsch (for the ALICE \fnref{col1} Collaboration)}
\fntext[col1] {A list of members of the ALICE  Collaboration and acknowledgements can be found at the end of this issue.}
\address{CERN, 1211 Geneva 23, Switzerland}

\begin{abstract} 
Full jet reconstruction in ALICE uses the combined information from charged and neutral particles. 
Essentially all jet constituents can be  measured with large efficiency down to very low transverse momenta ($\pt > 150 \, \mmom$). 
This has the advantage to introduce a minimum bias on the jet fragmentation, 
in particular for low jet momenta and in the presence of quenching.
In this article, we present preliminary results from reconstruction of charged jets 
in Pb--Pb collisions at $\sqrt{s_{\rm NN}} = 2.76 \, \tev$.  The inclusive charged jet spectrum, 
the jet nuclear modification factors ($R_{\rm AA}$, $R_{\rm CP}$), 
the ratio of spectra measured with different resolution parameters and hadron-jet correlations are discussed. 
For pp data at the same center of mass energy, 
the inclusive spectrum of fully reconstructed jets and its resolution parameter dependence are reported.
\end{abstract} 

\end{frontmatter} 


\section{Introduction}
The analysis of hadronic jets in heavy-ion collisions represents a formidable tool to study the properties of the Quark-Gluon Plasma.
Jets emerge from high-\pt\ quarks and gluons produced in hard scatterings during the very early phase of the reaction. 
The partons traverse the medium losing energy through elastic scattering and gluon radiation, 
a process in general called jet quenching \cite{Wiedemann:2009}.
Comparing heavy ion collisions to more elementary collisions like pp
the prominent experimentally observable effects of jet quenching 
are the decrease of the jet yield, energy imbalance of di-jet events, 
and the modification of the fragmentation function and the angular distribution of energy with respect to the jet axis.

At the LHC, rates are high at transverse energies where jets can be reconstructed above 
the fluctuations of the background energy contribution from the underlying event.
In particular, for jet transverse energies  $\et > 100 \, \gev$ the influence of the underlying 
event is relatively small allowing for robust jet measurements \cite{ATLAS, CMS}.
However, the measurements of the suppression of single particle production \RAA\
show that quenching effects are strongest for intermediate transverse momenta ($\RAA < 0.2$ for $4 < \pt < 20 \, \gmom $ corresponding 
to parton \pt\ in the range $\approx 6- 30 \, \gmom$) \cite{raa}.
The objective of ALICE is to access this \pt\ region introducing the smallest possible bias on the jet fragmentation by 
measuring jet fragments down to low \pt\ ($> 150 \, \mmom$).  
The key elements of this analysis are the detailed measurement of the  underlying event fluctuations (published in Ref. \cite{Abelev:2012ej}),  
the development of robust deconvolution procedures for spectra measured with low constituent \pt\ cuts \cite{hp_marta} 
and the suppression of fake jets (random combination of uncorrelated particles) at low \pt\  \cite{hp_gabriel}. 

\begin{figure}[ht]
\begin{minipage}[b]{0.5\linewidth}
\centering
\includegraphics[width=1.\textwidth]{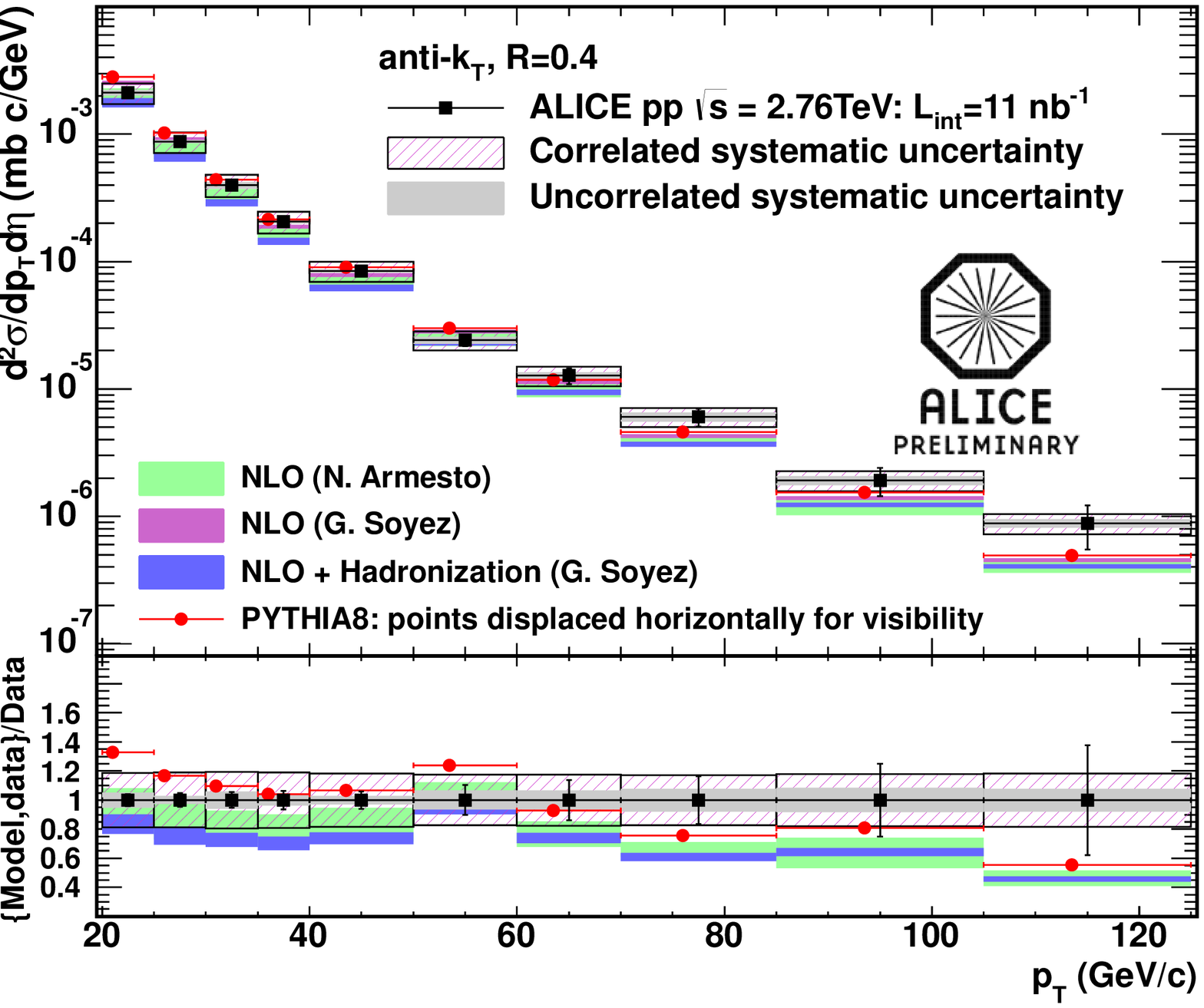} 
\caption{Differential cross section of fully reconstructed jets in pp collisions at $\sqrt{s}
=$~2.76~TeV compared to NLO pQCD \cite{Soyez, Armesto} and PYTHIA8 \cite{Sjostrand}.}
\label{fig:jets_full_pp}
\end{minipage}
\hspace{0.1cm}
\begin{minipage}[b]{0.49\linewidth}
\centering
\includegraphics[width=1.0\textwidth]{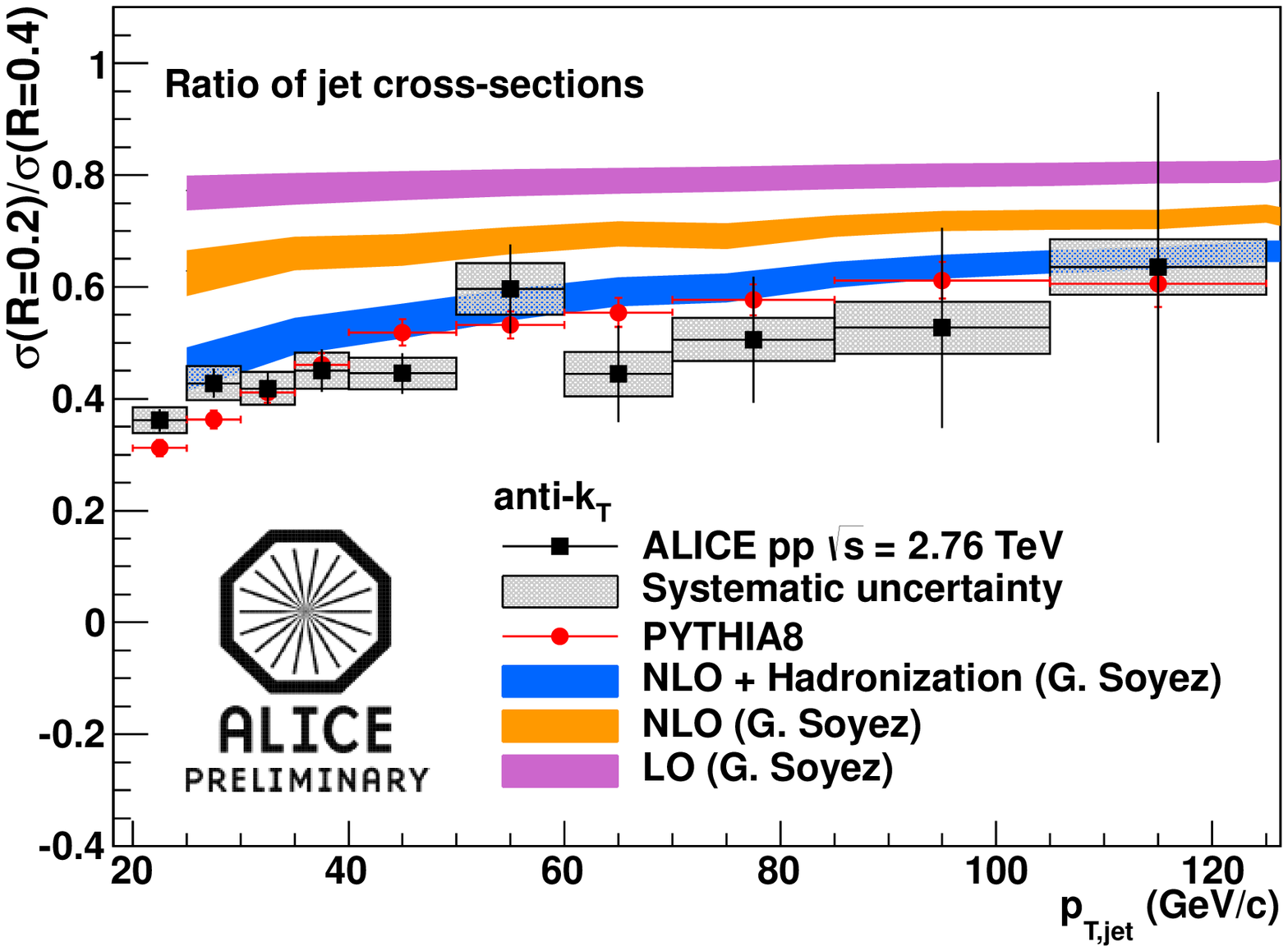} 
\caption{Ratio of jet cross sections reconstructed with $R = 0.2$ and $R=0.4$ in pp collisions at $\sqrt{s}
= $~2.76~TeV compared NLO pQCD \cite{Soyez, Armesto} and PYTHIA8 \cite{Sjostrand}.}
\label{fig:jets_ratio_pp}
\end{minipage}
\end{figure}

\section{Jet Reconstruction in ALICE}
In ALICE, full jet reconstruction uses the combined information from charged and neutral particle measurements. 
Charged particle momentum vectors are measured with the central tracking detectors, 
the Time Projection Chamber (TPC) and the Inner Tracking System (ITS) covering the full azimuth and $|\eta| < 0.9$.
Energy and direction of neutral particles are measured with the Pb-scintillator sampling ElectroMagnetic Calorimeter (EMCal), 
covering $1/3$ of the azimuth and $|\eta| < 0.7$. A detailed description of the ALICE experiment is given in Ref. \cite{alice}.

For jet reconstruction the anti-\kt\ algorithm from the FastJet package \cite{Cacciari:2011ma} 
with resolution parameters $R$ varying between 0.2 and 0.4 is used. 
The jet 4-momentum vector is calculated using the boost invariant \pt\ recombination scheme.
Analysis with charged jets using only tracking information and fully reconstructed jets including EMCal information have been performed.
The input for charged jets are tracks with $\pt > 150 \, \mmom$. 
The input for fully reconstructed jets is the charged jet input adding the EMCal cluster energy with \et\ above 
$150\, \mmom$ after correcting for charged particle energy contributions. 
These jets are required to be fully contained in the EMCal acceptance.

Jet-by-jet we correct for the energy contribution from charged particles to the energy measured with EMCAL and the contribution from the underlying event.
The sum of momenta of charged tracks matching the EMCal clusters from the cluster energy is subtracted resetting negative values to zero:
\begin{equation}
E^{\rm  corr}_{\rm clus} =  E^{\rm raw}_{\rm clus} - c \cdot \sum{p^{\rm matched}}; \, \, E^{\rm  corr}_{\rm clus} > 0
\end{equation}
The nominal value of $c$ is unity and is varied over a wide range to estimate the systematic uncertainty of this
procedure.

In \PbPb\ collisions, one has also to subtract the contribution of the Underlying Event (UE) from the reconstructed jet $\pt$. 
The summed \pt\ from the background is calculated as the product of mean momentum density $\rho$ and the jet area $A^{\rm Jet}$,  
where $\rho$ is determined using the \kt\ -algorithm \cite{Cacciari:2007fd} via:
$\rho = {\rm median} ( {p_{\rm T}^{\rm Jet} \over A^{\rm Jet}} )$.
Further corrections can only be applied on the raw spectrum bin-by-bin or using unfolding techniques. These corrections
comprise the unmeasured \pt\ from neutrons and ${\rm K}^{\rm 0}_{\rm L}$, 
as well as the tracking efficiency and the corresponding jet-by-jet fluctuations of these quantities.
In \PbPb\ collisions, we also correct for the smearing of the spectra induced by the UE energy fluctuations.
This smearing is quantified by $\delta \pt $, the difference between
the UE corrected summed \pt\ and the true jet \pt : 
$\delta{p_{\rm T}} = (p_{\rm  T}^{\rm  rec} - \rho A^{\rm Jet}) - p_{\rm T}^{\rm true}$.

A data-driven method to determine the distribution of $\delta \pt $
consists of embedding different objects into measured \PbPb\
collisions \cite{Abelev:2012ej}.
These objects can be single high-\pt\ tracks, jets or random cones. An advantage of the anti-\kt\ algorithm is, that the distribution does 
not depend significantly on the embedded object. The distribution is almost Gaussian with enhanced tails towards positive differences, owing to the 
pile-up of jets in the same jet area.
For a resolution parameter $R = 0.2$ the width ($\sigma$) of the Gaussian amounts to $6.2 \, \gmom$ ($4.5 \, \gmom$) summing neutral and charged \pt\ 
(for charged particles only).

\begin{figure}[ht]
\begin{minipage}[b]{0.5\linewidth}
\centering
\includegraphics[width=1.00\textwidth]{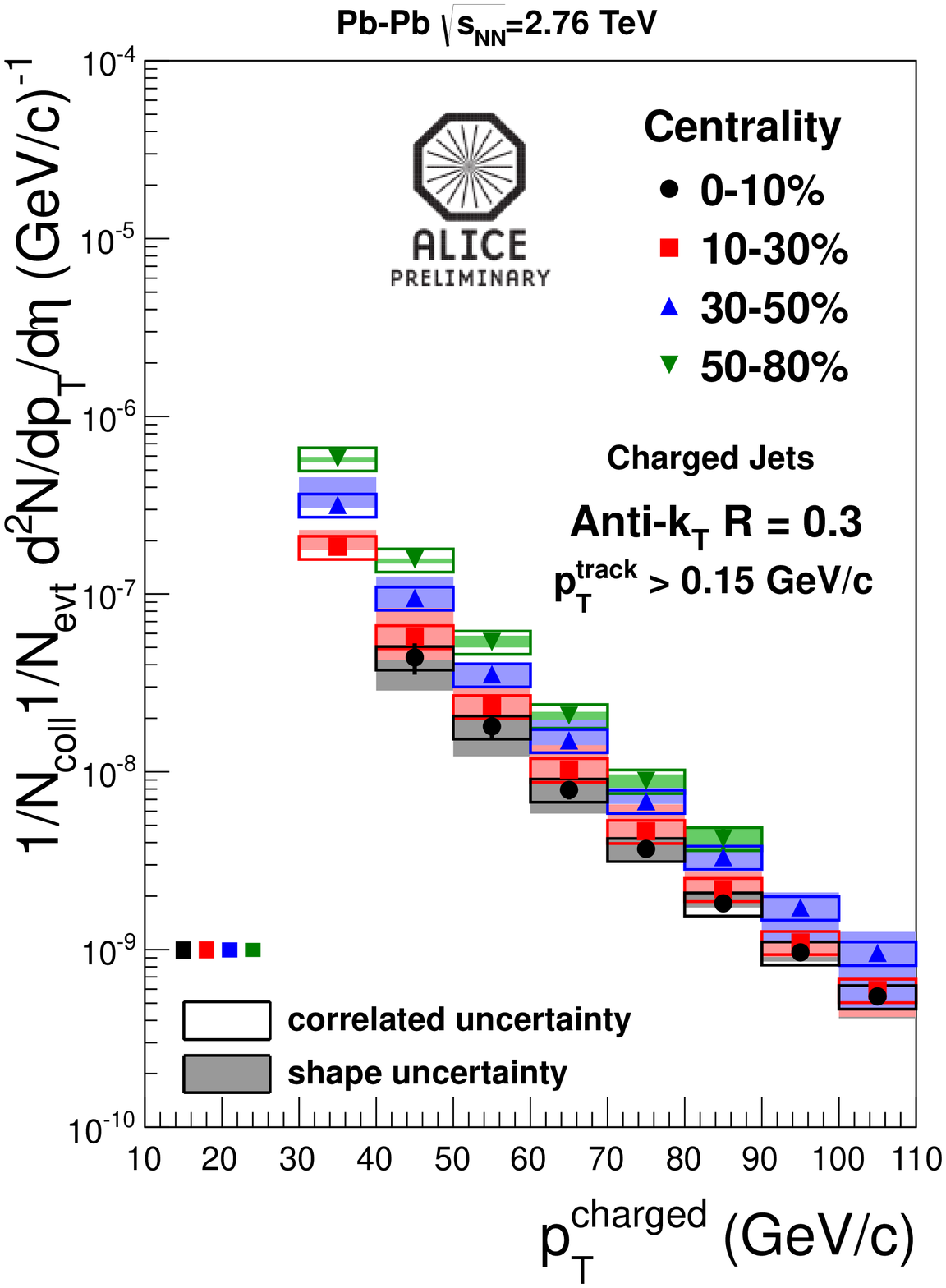} 
\end{minipage}
\begin{minipage}[b]{0.5\linewidth}
\centering
\includegraphics[width=0.95\textwidth]{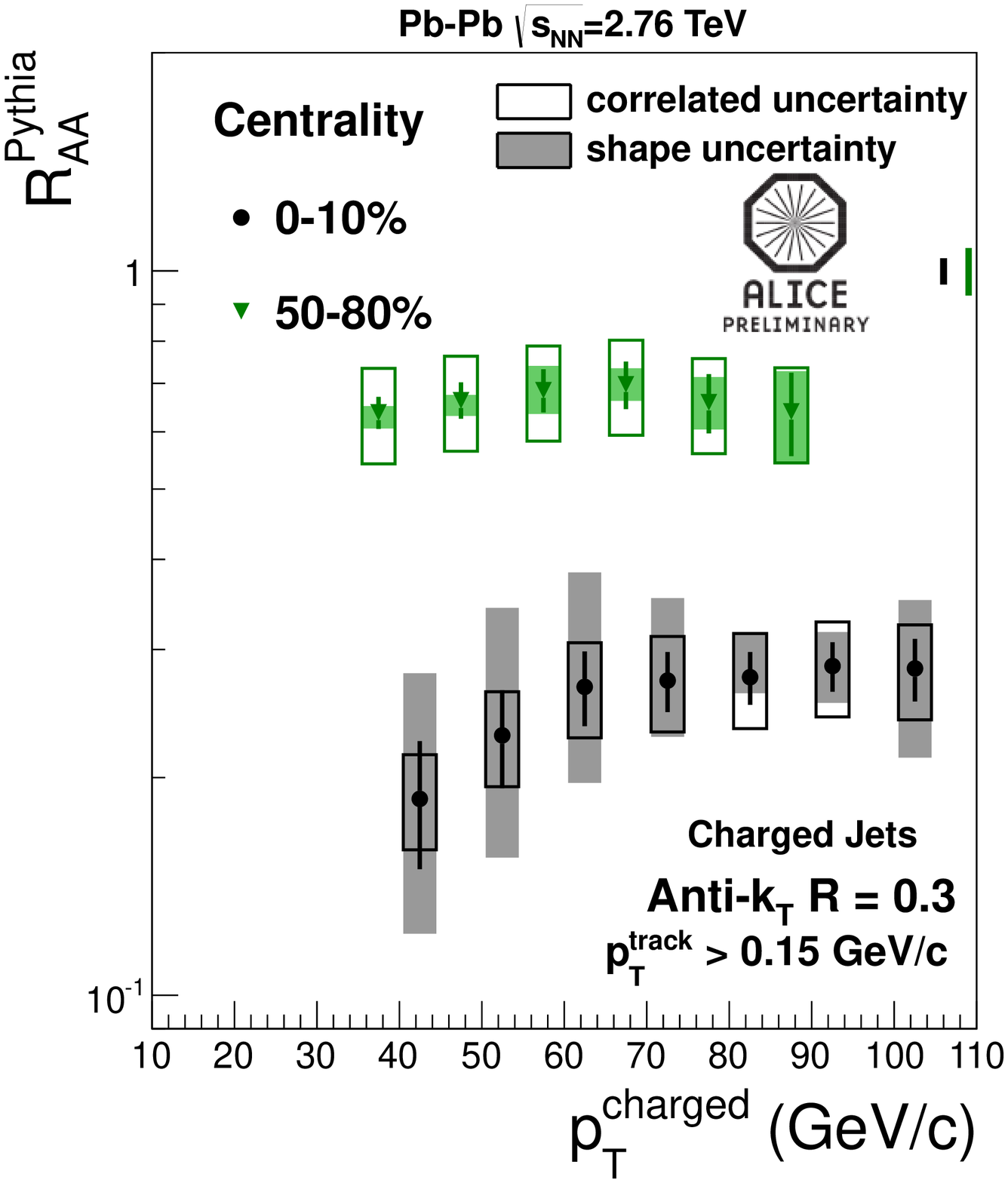} 
\end{minipage}
\caption{Left: Corrected jet spectrum with charged tracks for jet radius $R=0.3$.
Right: Nuclear modification factor for charged jets reconstructed in central and peripheral \PbPb\ collisions.
The pp reference is obtained from PYTHIA simulation at the same $\sqrt{s}$.}
\label{fig:jets_raa_pbpb}
\end{figure}

\section{Jet Spectra in pp at $\sqrt{s} = 2.76 \, \tev$}
In March 2011, a short pp reference run at the Pb--Pb energy of $2.76 \, \tev $ was taken
integrating $20 \, {\rm nb}^{-1}$ of rare triggers. This was the first running period in which the EMCal was 
fully installed allowing us to perform measurements with fully reconstructed jets up to $120 \, \gmom$ from 
 $11 \, {\rm nb}^{-1}$ of these data.
Fig. \ref{fig:jets_full_pp} shows the jet \pt\ spectrum reconstructed with a resolution parameter of $R = 0.4$. 
The jet energy scale uncertainty amounts to 4\% and it is mainly due to uncertainties in the missing neutral energy, 
the tracking efficiency and energy double-counting.
The jet \pt\ resolution $\Delta \pt / \pt$ amounts to 20\% and is dominated by the jet energy scale fluctuations, 
tracking and EMCal resolutions.
Efficiency and resolution effects on the jet spectrum were taken into account by applying a bin-by-bin correction.
The measured spectrum is compared to NLO pQCD calculations \cite{Soyez, Armesto} and PYTHIA8 \cite{Sjostrand}. A good agreement 
is observed.
The measurement itself represents an important reference for our Pb--Pb measurements.

Ratios of differential cross-sections measured with different $R$ provide information on the energy distribution within the jet area (jet shape). 
Note that for jets reconstructed with the anti-\kt\ algorithm, $R$ is to a good approximation the radius of a circular jet area.
Fig. \ref{fig:jets_ratio_pp} shows the measured ratio for $R = 0.2$ and $0.4$. 
Due to the fact that higher \pt\ jets are more collimated, the ratio rises with $\pt^{\rm Jet}$; in agreement with the pQCD and PYTHIA8
calculations.

\begin{figure}
\centering
\includegraphics[width=0.6\textwidth]{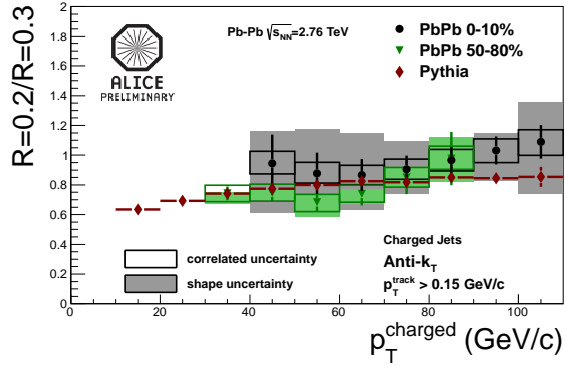} 
\caption{Ratio of reconstructed charged jets with different cone radii
  in central and peripheral \PbPb\ collisions at $\sqrt{s_{\rm NN}} = 
$~2.76 TeV compared to  PYTHIA (unquenched).}
\label{fig:jets_ratio_pbpb}
\end{figure}

\section{Jet Suppression in Pb--Pb}

Measurements of jet spectra in Pb--Pb collisions were obtained from a sample of $3 \cdot 10^7$ minimum bias events 
at  $\sqrt{s_{\rm NN}} = 2.76 \, \tev$ collected in November 2010. 
Since the electromagnetic calorimeter was only fully installed in the beginning of 2011, jet reconstruction 
was performed using charged particle information only.

Fig. \ref{fig:jets_raa_pbpb}~(left) shows the charged jet yields as a function of \pt\
normalized by the number of collisions (resolution parameters $R = 0.3$) for four centrality classes.
The influence of detector effects and 
background fluctuations were corrected for by applying a regularized unfolding procedure with $\chi^2$ minimization.
The systematic uncertainties are dominated by the choice of the unfolding parameters ($4\%$) and the jet energy scale 
corrections ($4-10\%$).
 
Comparing the spectra for different  centrality classes a 
sizable suppression increasing with centrality can be observed. 
In order to study the modifications of the Pb--Pb spectra with respect to an incoherent 
superposition of binary nucleon--nucleon collisions,  the nuclear modification factor $R_{\rm AA}^{\rm Jet}$ was calculated and is shown in Fig. \ref{fig:jets_raa_pbpb}~(right).
As a reference we use the spectra from pp collisions at the same centre-of-mass energy simulated by the PYTHIA MC \cite{Sjostrand}.
The results for the highest centrality bin (0-10\%) and the lowest one (50-80\%) for $R=0.2$ are shown.
A strong nuclear suppression qualitatively and quantitatively similar to the $\RAA$ of inclusive hadrons is observed in the most central 
collisions.

A possible effect of jet quenching is the redistribution of the radiated energy within the jet cone leading to jet shape modifications.
As mentioned earlier the ratios of differential cross-sections measured with different $R$ can provide information on these jet shape modifications.
Fig. \ref{fig:jets_ratio_pbpb} shows the measured ratio ($R = 0.2$)/($R = 0.3$). No modifications of the jets measured in central Pb--Pb 
collisions with respect to more peripheral collisions or the PYTHIA pp reference are observed within the present large experimental uncertainties.

\begin{figure}[ht]
\begin{minipage}[b]{0.5\linewidth}
\centering
\includegraphics[width=1.00\textwidth]{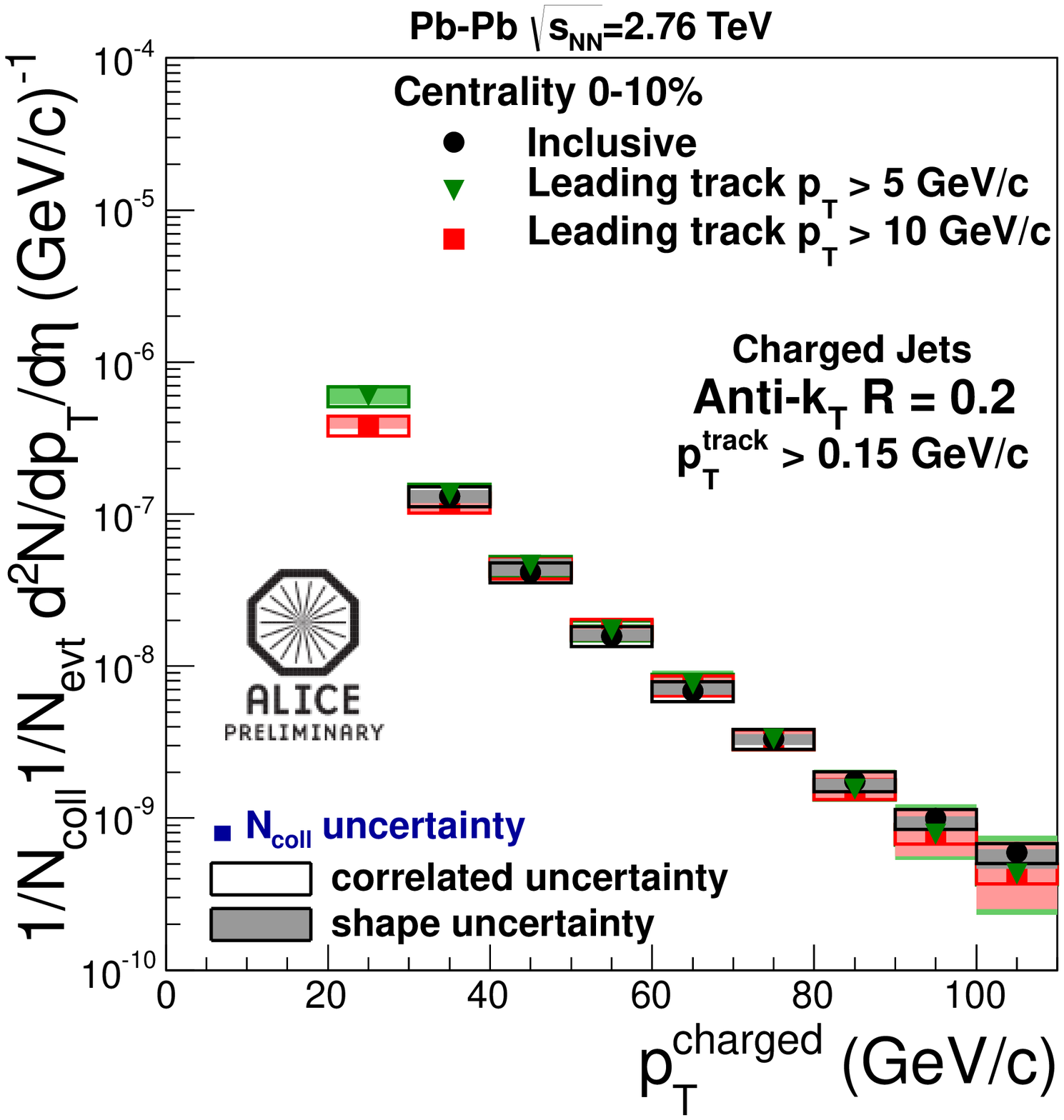} 
\end{minipage}
\begin{minipage}[b]{0.5\linewidth}
\centering
\includegraphics[width=1.0\textwidth]{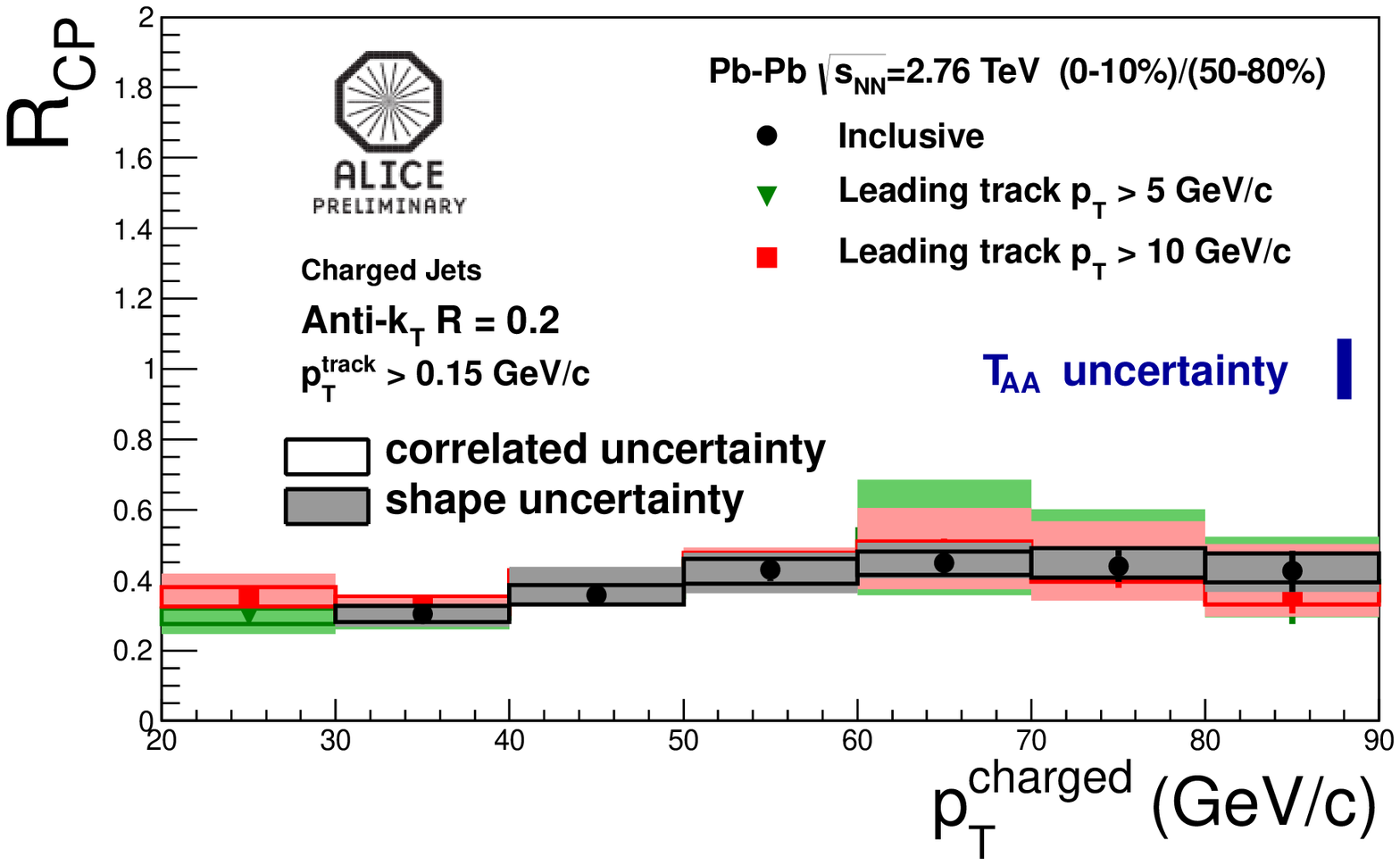} 
\end{minipage}
\caption{Charged jet measurements in Pb--Pb using the  high-\pt\ hadron tag and resolution parameter $R=0.2$. 
Left: Corrected charged jet spectrum for central Pb-Pb collisions.
Right: $R_{\rm CP}$ relative to the centrality bin $50-80 \%$.
}
\label{fig:jets_leading}
\end{figure}

The requirement of a high $\pT$ leading hadron within  the jet area reduces the contribution from combinatorial (fake) jets.
This leads to an improved stability of the unfolding and allows us to assess lower $p_{\rm T, Jet}^{ch}$. Using a resolution parameter 
$R = 0.2$ we can measure the charged jet spectrum down to $20 \, \gmom$.
In Fig. \ref{fig:jets_leading}~(left) 
we compare the inclusive spectrum to those with $p_{T}^{\rm leading} > 5 \, \gmom $ and  $p_{T}^{\rm leading} > 10 \, \gmom $ requirement.
For central collisions (0-10\%), we observe no change of fragmentation within uncertainties except for the lowest $p_{\rm T, Jet}^{ch}$ bin. 
To quantify the jet suppression in central collisions we plot in Fig. \ref{fig:jets_leading}~(right) the ratio of the jet yield with respect to centrality 
$50-80 \%$  ($R_{\rm CP}$). It shows a suppression pattern similar to \RAA .
For more details see the contribution of M. Verweij to these proceedings.

\section{Hadron-Jet Correlations}
In the presence of strong partonic energy loss, leading hadron emission is biased towards the surface of the reaction zone.
Hence, the in-medium path-length of the parton produced back-to-back to the leading particle is above average and the
jet emerging from this parton can be expected to show stronger than average suppression and structure modifications \cite{Renk}.
To exploit this effect we study the conditional yield of jets produced opposite in azimuth to a leading particle
within a given \pt\ range $p_{T}^{\rm min} -  p_{T}^{\rm max}$:

\begin{equation}
Y(p_{T, \rm Jet}^{\rm ch};p_{T}^{\rm min}, p_{T}^{\rm max}) = 
{{1} \over {N_{\rm tr}}} {{{\rm d}N(p_{T, \rm Jet}^{\rm ch};p_{T}^{\rm min}, p_{T}^{\rm max})} \over {{\rm d}p_{T, \rm Jet}^{\rm ch}}}
\end{equation}
The method has additional advantages.
There is no bias on the fragmentation of the recoiling jet.
Moreover, the requirement of a correlated high-\pt\ hadron tags hard scatterings 
and suppresses the combinatorial, fake jet background at low jet $\pt$.

\begin{figure}
\centering
\includegraphics[width=0.6\textwidth]{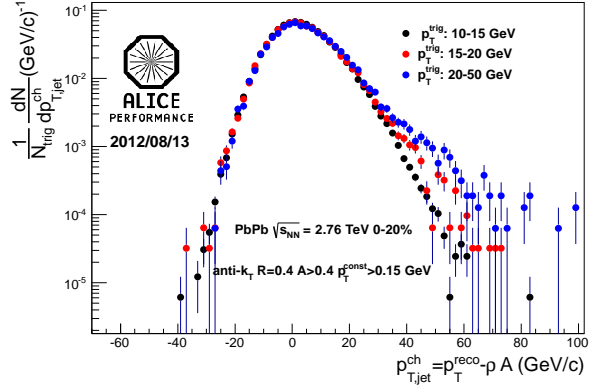} 
\caption{
Uncorrected recoil jet yield per trigger differential in $\pt$ for three trigger \pt\ ranges.
}
\label{fig:jets_recoil_pbpb}
\end{figure}

Fig. \ref{fig:jets_recoil_pbpb} shows the uncorrected recoil charged jet yield per trigger differential in $\pt$.
The azimuthal distance between trigger hadron and jet is required to be in the region $\pi \pm 0.6$.
Three trigger \pt\ bins are shown.
The low jet \pt\ region is almost independent of the trigger \pt\  and dominated by fake and uncorrelated jets.
The high jet \pt\ region shows a clear hardening with increasing trigger \pt\ and is dominated by high $Q^2$ scattering.

The uncorrelated background does not depend on the trigger hadron \pt\ range. 
Hence, in the difference between the per trigger yields for two different trigger \pt\ ranges 
$\Delta_{\rm recoil}(p_{T, \rm Jet}^{\rm ch})$
the uncorrelated background cancels. We define:
\begin{equation}
\Delta_{\rm recoil}(p_{T, \rm Jet}^{\rm ch})= 
Y(p_{T, \rm Jet}^{\rm ch};p_{T}^{\rm min}, p_{T}^{\rm max}) -
Y(p_{T, \rm Jet}^{\rm ch};p_{T, ref}^{\rm min}, p_{T, ref}^{\rm max})
\end{equation}

Modifications of the conditional yields in central Pb--Pb collisions have been studied measuring the ratio
\begin{equation}
\Delta I_{\rm AA}(p_{T, Jet}^{\rm ch})= 
\Delta_{\rm recoil}^{\rm Pb - Pb} / \Delta_{\rm recoil}^{\rm pp}
\end{equation}

\begin{figure}
\centering
\includegraphics[width=0.49\textwidth]{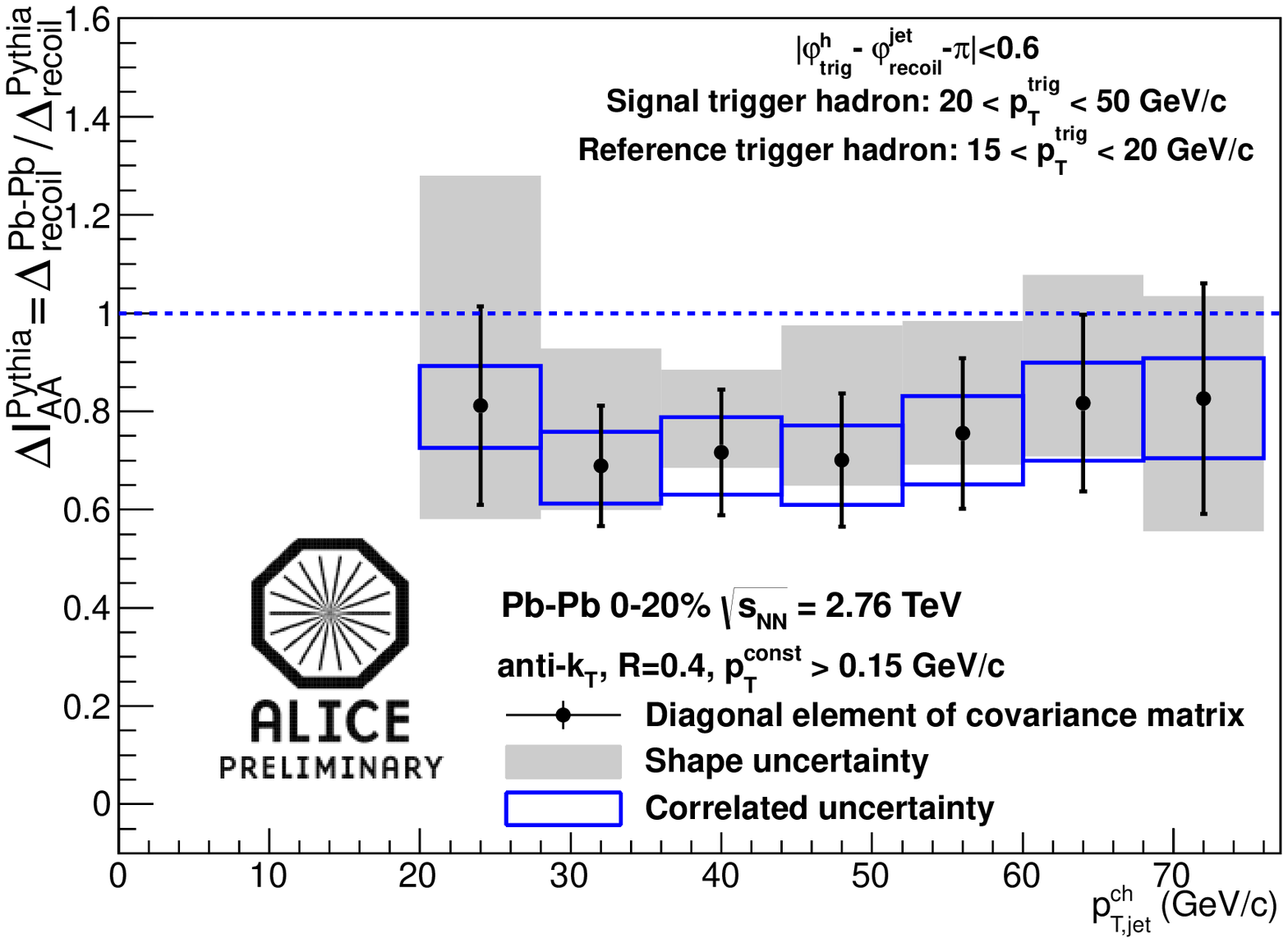} 
\includegraphics[width=0.49\textwidth]{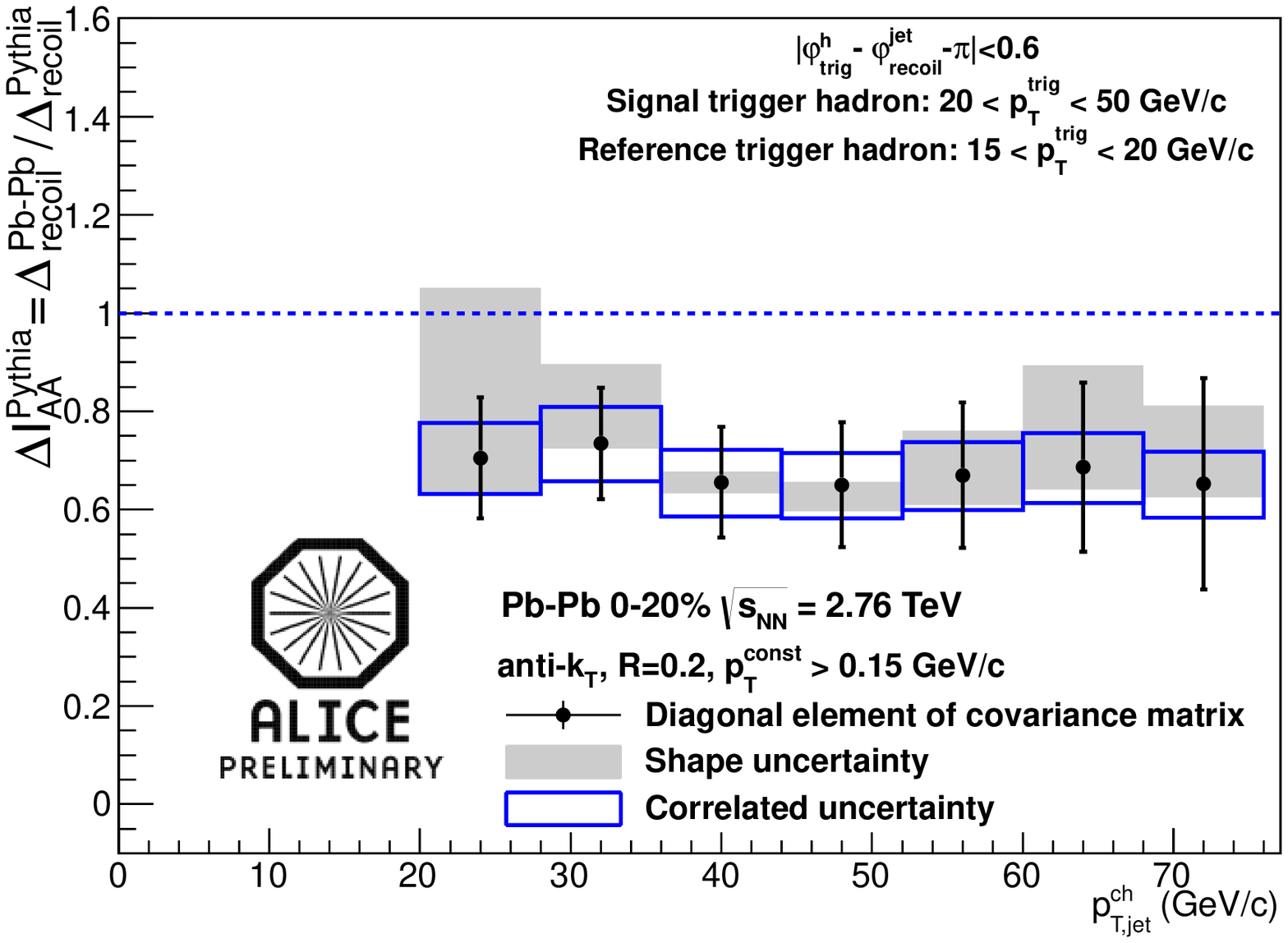} 
\caption{The ratio $\Delta I_{\rm AA}$ using PYTHIA as the pp reference for two different resolution parameter:
$R=0.4$~(left) and $R=0.2$~(right).
}
\label{fig:jets_hadron-jet_pbpb}
\end{figure}

The fully corrected ratios for two different resolution parameters are shown in Fig. \ref{fig:jets_hadron-jet_pbpb}. 
A trigger \pt\ range $20-50 \, \gmom$ was used for the signal and $15-20 \, \gmom $ for the reference.
Again results of a PYTHIA simulation were used as pp reference.
The ratio is close to unity  close to the conditional away-side hadron-hadron yield suppression measured by ALICE at a lower $Q^2$
\cite{IAA}.
Note, however, that the hadron-jet pair suppression is strong. It is approximately
$\RAA (20 \, \gmom ) \times \Delta I_{\rm AA} \approx 0.25 \times 0.75 \approx 0.2$.
For more details see the contribution of L. Cunqueiro to these proceedings.

\section{Isolated Photon-Hadron Correlations in pp}
A very attractive possibility for studying jets at lower energy in heavy-ion collisions consists of
tagging jets by $\gamma$-jet correlations. Direct gammas result in leading order from annihilation and Compton scattering 
($q +g(q) → \gamma +q(g)$).
In contrast to partons, gammas penetrate the medium almost unaltered. They approximately
balance the transverse energy of the original parton and, hence, can be used to assess the jet
energy.
Experimentally one can tag such processes by identifying leading isolated photons and correlated associated
charged hadrons in the opposite azimuthal direction. 
Photons are detected in the EMCal. Photon candidates are reconstructed from clusters of energy deposited in the \pt\ range 
from 8 to $25 \, \gmom$. These candidates are dominated by background from electromagnetic decays of neutral mesons. 
Since these particles are mainly produced in jets, the background 
can be reduced by imposing an isolation criteria. In our analysis we require that no particle with $\pt > 0.5 \, \gmom$ 
is found within a cone of radius $R = 0.4$.

\begin{figure}
\centering
\includegraphics[width=0.5\textwidth]{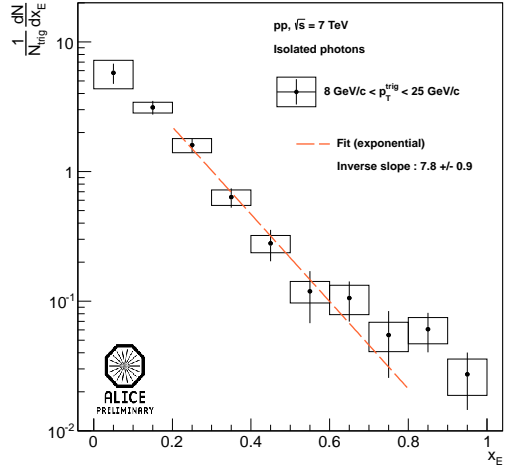} 
\caption{$x_{E}$ distributions (with : $x_{E} = - \pt ^{\rm asso} /\pt ^{\rm trig} \cos{\Delta \varphi}$) 
of isolated photons for the \pt\ range $\pt =  8 -25 \, \gmom $. The red line show the result of an 
exponential fit in the range [0.2-0.8].
}
\label{fig:gamma-hadron}
\end{figure}

The fragmentation function of the recoiling jet can be approximated by the $x_{E}$ distribution, 
where $x_{E} = p_{T}^{hadron}/  p_{T}^{\gamma} \cos{\Delta \varphi} $.
Fig. \ref{fig:gamma-hadron} shows the measured  $x_{E}$ distribution for $p_{T}^{hadron} > 0.2 \, \gmom$ after correction 
for residual background. The distribution can be fit by an exponential function with inverse slope $7.8 \pm 0.9$.
The measurement of the $x_{E}$ distribution and the inverse slope from $\gamma$-hadron correlations represents a first
important step towards corresponding measurements in Pb--Pb.
For more details see the contribution of N. Arbor to these proceedings.

\section{Summary}
Preliminary measuremenst in central Pb-Pb collisions at 
$\sqrt{s_{\rm NN}} = 2.76 \, \tev$ show a strong suppression of the inclusive charged jet yield.
Using PYTHIA as a reference , the nuclear suppression factor \RAA\ for charged jets amounts to $0.2-0.3$ in the range 
$30 < p_{T, \rm ch}^{\rm ch} < 100 \, \gmom $, lower than the  inclusive hadron \RAA\ at similar parton \pt .
No indication of energy redistribution within experimental uncertainties is observed from ratios of jet yields ($R=0.2$)/($R=0.3$).
The conditional hadron-jet yield is suppressed by a factor of 0.75 with respect to the PYTHIA reference. 
This is close to the conditional away-side hadron-hadron yield suppression measured by ALICE at a lower $Q^2$.
The yield and conditional yield suppression patterns are qualitatively and to some extent quantitatively similar 
for single hadrons and jets.
This is qualitatively consistent with the expectations for partonic energy loss through radiation mainly outside the jet cone
and in-vacuum fragmentation of the remnant parton.
First preliminary results from $\gamma $-hadron correlations measurements and $x_{\rm E}$ in pp at $\sqrt{s}= 7\,  \tev$
have been presented. They represent an important step towards a corresponding measurement in Pb--Pb.


\section*{References}

\begin{thebibliography}{20} 
\bibitem{Wiedemann:2009} U.A.~Wiedemann, {\em Jet Quenching in Heavy Ion Collisions}, [arXiv:0908.2306].
\bibitem{ATLAS} G. Aad {\it al.} (ATLAS Collaboration), Phys. Rev. Lett. {\bf 105}, 252303 (2010).
\bibitem{CMS} S. Chatrchyan {\it et al.} (CMS Collaboration), Phys. Rev. {\bf C84}, 024906 (2011).
\bibitem{raa} K. Aamodth {\it et al.} (ALICE Collaboration), Phys. Lett. {\bf B696} 30-39 (2011).
\bibitem{Abelev:2012ej} B.~Abelev {\it et~al.} (ALICE Collaboration), JHEP {\bf 1203}, 053 (2012).
\bibitem{hp_marta} M. Verweij, (ALICE Collaboration), {\em Measurement of jet spectra in Pb-Pb collisions at $\sqrt{s_{NN}}$=2.76 TeV with the ALICE detector at the LHC  }, [arXiv:1208.6169]. 
\bibitem{hp_gabriel} G.O.V.~de~Barros {\it et al.},  {\em Data-driven analysis methods for the measurement of reconstructed jets in heavy ion collisions at RHIC and LHC}, [arXiv:1208.1518].
\bibitem{alice} K. Aamodt et al. (ALICE Collaboration), JINST 3, S08002 (2008).
\bibitem{Cacciari:2011ma} M.~Cacciari, G.~P. Salam and G.~Soyez, Eur.Phys.J. {\bf C72}, 1896 (2012).
\bibitem{Cacciari:2007fd} M.~Cacciari and G.~P. Salam,Phys. Lett. {\bf B659}, 119 (2008).
\bibitem{Soyez} G. Soyez, Phys. Lett. B 698, 59 (2011) [arXiv:1101.2665v1]; private communication.
\bibitem{Armesto} 
N. Armesto, private communication. Calculations based on:\\
S. Frixione, Z. Kunszt, and A. Signer, Nucl. Phys. {\bf B467}, 399-442 (1996);\\
S. Frixione, Nucl. Phys. {\bf B507}, 295-314 (1997).
\bibitem{Sjostrand} T.~Sj{\"o}strand, S.~Mrenna and P.~Skands, JHEP {\bf 05}, 026 (2006);\\
T.~Sjostrand, S.~Mrenna and P.~Z. Skands, Comput. Phys. Commun. {\bf 178}, 852 (2008).
\bibitem{Renk} T~Renk, [arXiv:1204.5572].
\bibitem{IAA} K. Aamodt {\it et. al.} (ALICE Collaboration),   Phys. Rev. Lett. {\bf 108}, 092301 (2012).
\end{thebibliography}
\end{document}